\documentclass[aps,showpacs,preprint,superscriptaddress]{revtex4}
\usepackage{graphicx}
\usepackage{subfigure}
\usepackage{float}
\usepackage{amsmath}
\usepackage{amsfonts}
\usepackage{bm}
\usepackage{bbm}
\usepackage{txfonts}
\usepackage{array}
\usepackage{amssymb}
\usepackage{color}
\begin{document}
\title{Enhanced pair production in frequency modulated Sauter potential wells}
\author{Li Wang}
\affiliation{Key Laboratory of Beam Technology of the Ministry of Education, and College of Nuclear Science and Technology, Beijing Normal University, Beijing 100875, China}
\author{Binbing Wu}
\affiliation{Key Laboratory of Beam Technology of the Ministry of Education, and College of Nuclear Science and Technology, Beijing Normal University, Beijing 100875, China}
\author{Lie-Juan Li}
\affiliation{Key Laboratory of Beam Technology of the Ministry of Education, and College of Nuclear Science and Technology, Beijing Normal University, Beijing 100875, China}
\author{B. S. Xie  \footnote{Corresponding author. Email address: bsxie@bnu.edu.cn}}
\affiliation{Key Laboratory of Beam Technology of the Ministry of Education, and College of Nuclear Science and Technology, Beijing Normal University, Beijing 100875, China}
\affiliation{Beijing Radiation Center, Beijing 100875, China}

\date{\today}

\begin{abstract}
  Electron-positron pair production in frequency modulated Sauter potential wells is investigated in the framework of the computational quantum field theory. In combined potential wells with a static Sauter potential well and a frequency modulated oscillating one, the modulated amplitude has a large effect on the number of created pairs. The optimal modulation amplitude of frequency at different center frequencies is obtained, which increases the number of electrons at about two times. However, for a single oscillating potential well with frequency modulation, chirp effect is sensitive to the center frequency, and the number of electrons can be enhanced even to four orders of magnitude at a regime of low center frequency. It implies that for a slowly oscillating Sauter potential well, the chirp effect through the frequency modulation is better than adding a static potential well to improve the pair production.
\end{abstract}

\maketitle
\section{Introduction}

Since the positron was proposed by Dirac to explain the negative solution of Dirac equation, the generation of electron-positron pairs in vacuum with strong electric field aroused great concern \cite{Dirac1928}. Soon after that the existence of positrons was proved experimentally \cite{Anderson1933}, many theoretical studies were shown to explain the mechanism to produce positrons \cite{Sauter1931,Heisenberg1936,Schwinger1951,Brezin1970,Marinov1977}. So far,
several methods were developed, such as the worldline instanton technique \cite{Gies2005,Dunne2006,Ilderton2015,Schneider2016}, the Dirac-Heisenberg-Wigner formalism \cite{Li2015,Blinne2016,Kohlfurst2018,Olugh2019}, the quantum Vlasov equation solution method \cite{Kluger1991,Alkofer2001,JiangMin2013,Li2014,Nuriman2012,Sitiwaldi2017,Xie2017}, the computational quantum field theory \cite{Krekora2005,Lv2013,Tang2013,Jiang2013,Jiang2014,Liu2014,Wang2016,Gong2018} and so on.
Overall, there are two mechanisms proposed.
The Schwinger mechanism is due to the quantum tunneling effect and requires the electric field to reach $10^{16}\rm{V/cm}$, which is beyond the current experimental conditions \cite{Schwinger1951}. The multi-photon process is due to a transition between the negative and positive states through photon absorption, which depends on the frequency of the alternating field \cite{Brezin1970,Marinov1977}. Due to the fact that current ultrafast laser can not produce a considerable number of positrons, people proposed new ways to reduce the threshold or increase the yield.

Taking both the static potential well and the alternating field into account, the pair creation in a symmetric potential well was investigated \cite{Tang2013}. It is found that the symmetric potential produces more electrons than the asymmetric potential, and combined potential wells are more favorable for the creation of pairs than a static one.
Due to the dynamically assisted Schwinger mechanism, the pair creation rate can also be dramatically enhanced by combining a strong and slowly varying electric field with a weak and rapidly changing one \cite{Li2014,Nuriman2012,Schutzhold2008}.
When the photon energy equals the distance between the bound state and the Dirac sea level, the pair creation can be enhanced, which is the bound state resonance enhanced mechanism \cite{Wu:2019weg}. For different bound states, different frequencies are needed to achieve resonance so that the frequency chirp field was also considered to enhance the creation of pairs \cite{Dumlu2010,Jiang2013,Abdukerim2017,Olugh2019}. Recently the process of pairs creation in a spatially homogeneous but frequency modulation (FM) electric field is studied \cite{Gong2019}, where the result shows that the number of pairs can be enhanced or weakened by adjusting the parameter of modulation frequency.

In this paper, we study the enhancement of electron-positron pairs in FM Sauter potential wells, and optimize the modulation parameters. Considering chirp effect, we study the pair creation process in combined potential wells composed of a static Sauter potential well and a FM one in the framework of the computational quantum field theory.
Moreover, the single oscillatory potential well is also studied. It is found that chirp effect has a great influence on the number of created pairs at low center frequencies, and the detailed explanation and analysis are given.

This paper is organized as follows. In Sec. II, the scheme is introduced briefly for the computational quantum field theory, which solves the Dirac equation in operator formalism and get the required number of created electron-positron pairs. In Sec.III, FM Sauter potential wells are presented, in which the time evolution of created electrons under different FM amplitudes and center frequencies are simulated. In Sec.IV, we summarize our work.

\section{Outline of computational quantum field theory}\label{section2}

Let us first describe the computational quantum field theory. The time evolution of the operator $\hat{\psi} (z,t)$ for the electron-positron yield in a potential $V(z,t)$ is given by the Dirac equation\cite{Krekora2005},
\begin{equation}\label{Eq Dirac}
i\partial \hat{\psi} \left(z,t\right) / \partial{t}=\left[c\alpha_z \hat{p}_z+\beta c^2+V\left(z,t\right)\right] \hat{\psi}\left(z,t\right),
\end{equation}
where $\alpha_z$ and $\beta$ are Dirac matrices, $c$ is the speed of light in vacuum, $\hat{p}_z$ is the momentum operator of $z$ direction, $V\left(z,t\right)$ is external field that varies with time $t$ in the $z$ direction. We use the atomic units (a.u.) as $\hbar=e=m_e=1$.
By introducing the creation and annihilation operators, the field operator $\hat{\psi}(z,t)$ can be decomposed as follows:
\begin{equation}\label{Eq Field Operator}
\begin{aligned}
\hat{\psi}(z,t)&=\sum_{p}\hat{b}_p(t)W_p(z)+\sum_{n}\hat{d}_n^{\dag}(t)W_n(z) \\
&=\sum_{p}\hat{b}_pW_p(z,t)+\sum_n\hat{d}_n^\dag W_n(z,t),
\end{aligned}
\end{equation}
where $p$ and $n$ denote the momenta of positive and negative energy states, $\sum_{p(n)}$ represents summation over all states with positive $($negative$)$ energy, $W_p(z)=\langle z|p\rangle(W_n(z)=\langle z|n\rangle)$ is field-free positive $($negative$)$ energy eigenstate. Note that $W_p(z,t)=\langle z|p(t)\rangle$ and $W_n(z,t)=\langle z|n(t)\rangle$ satisfy the single-particle time-dependent Dirac equation (\ref{Eq Dirac}).
From Eq.(\ref{Eq Field Operator}), we obtain
\begin{equation}\label{Eq fermion operators}
\begin{aligned}
\hat{b}_p(t)&=\sum_{p'}\hat{b}_{p'}U_{pp'}(t)+\sum_{n'}\hat{d}_{n'}^\dag U_{pn'}(t),\\
\hat{d}_n^\dag(t)&=\sum_{p'}\hat{b}_{p'}U_{np'}(t)+\sum_{n'}\hat{d}_{n'}^\dag U_{nn'}(t),\\
\hat{b}_p^\dag(t)&=\sum_{p'}\hat{b}_{p'}^\dag U_{pp'}^*(t)+\sum_{n'}\hat{d}_{n'}U_{pn'}^*(t),\\
\hat{d}_n(t)&=\sum_{p'}\hat{b}_{p'}^\dag U_{np'}^*(t)+\sum_{n'}\hat{d}_{n'}U_{nn'}^*(t),
\end{aligned}
\end{equation}
where $U_{pp'}(t)=\langle p|\hat{U}(t)|p'\rangle$, $U_{pn'}(t)=\langle p|\hat{U}(t)|n'\rangle$, $U_{nn'}(t)=\langle n|\hat{U}(t)|n'\rangle$, $U_{np'}(t)=\langle n|\hat{U}(t)|p'\rangle$ and the time-ordered propagator $\hat{U}(t)=\hat{T}\textrm{exp}\{{-i\int_0^t d\tau [c\alpha_z \hat{p}_z+\beta c^2+V(z,\tau)]}\}$.

In Eq.(\ref{Eq Field Operator}), the electronic portion of the field operator is defined as $\hat{\psi}_e(z,t)\equiv \sum_p \hat{b}_p(t)W_p(z)$. So we can obtain the probability density of created electrons by
\begin{equation}\label{Eq Density}
\begin{aligned}
\rho(z,t)&=\langle \text{vac}|\hat{\psi}_e^\dag (z,t)\hat{\psi}_e(z,t)|\text{vac}\rangle \\
&=\sum_n |\sum_p U_{pn}(t)W_p(z)|^2
\end{aligned}
\end{equation}
By integrating this expression over space, the number of created electrons can be obtained as
\begin{equation}\label{Eq Number}
N(t)=\int \rho(z,t)dz=\sum_p \sum_n |U_{pn}(t)|^2.
\end{equation}
The time-ordered propagator $U_{pn}(t)$ can be numerically calculated by employing the split-operator technique \cite{Tang2013}. Therefore, according to Eq.(\ref{Eq Density}) and Eq.(\ref{Eq Number}), we can compute various properties of the electrons produced under the action of the external potential.

\section{Numerical results}

In this work, we study combined potential wells with a static Sauter potential well and a FM one, which is given by
\begin{equation}\label{Eq Well}
\begin{aligned}
V(z,t)&=V_1 S(z) f(t)+V_2 \sin[\omega(t)t]S(z)\theta(t;t_0,t_0+t_1), \\
\end{aligned}
\end{equation}
where $S(z)=\{\tanh[(z-D/2)/W]-\tanh[(z+D/2)/W]\}/2$, $D$ is the width of the potential well, $W$ is the width of the potential edge. Here $V_1$ is the depth of the static potential well, $V_2$ is the amplitude of the oscillating potential well. The time dependent oscillation frequency is set to $\omega(t)=\omega_0+\Delta \omega \sin[\Omega (t-t_0)]$, where $\omega_0$ is the center frequency, $\Delta \omega$ is the FM amplitude, $\Omega$ reflects the speed of frequency change.
The function $f(t)=\sin[\pi t/2t_0]\theta(t;0,t_0)+\theta(t;t_0,t_0+t_1)+\cos[\pi (t-t_0-t_1)/2t_0]\theta(t;t_0+t_1,2t_0+t_1)$ describes the turning on and off processes of the potential well, and $\theta (t,t_1,t_2)$ is the step function. The static potential well is opened during $(0,t_0)$ and closed during $(t_0+t_1,2t_0+t_1)$. From $t_0$ to $t_0+t_1$, the oscillating potential well exists. In this paper, other parameters are set to $t_0=5/c^2$, $W=0.3\lambda_C$, and $\lambda_C=1/c$ is the Compton wavelength.

\begin{figure}[htbp]\suppressfloats
\includegraphics[width=12cm]{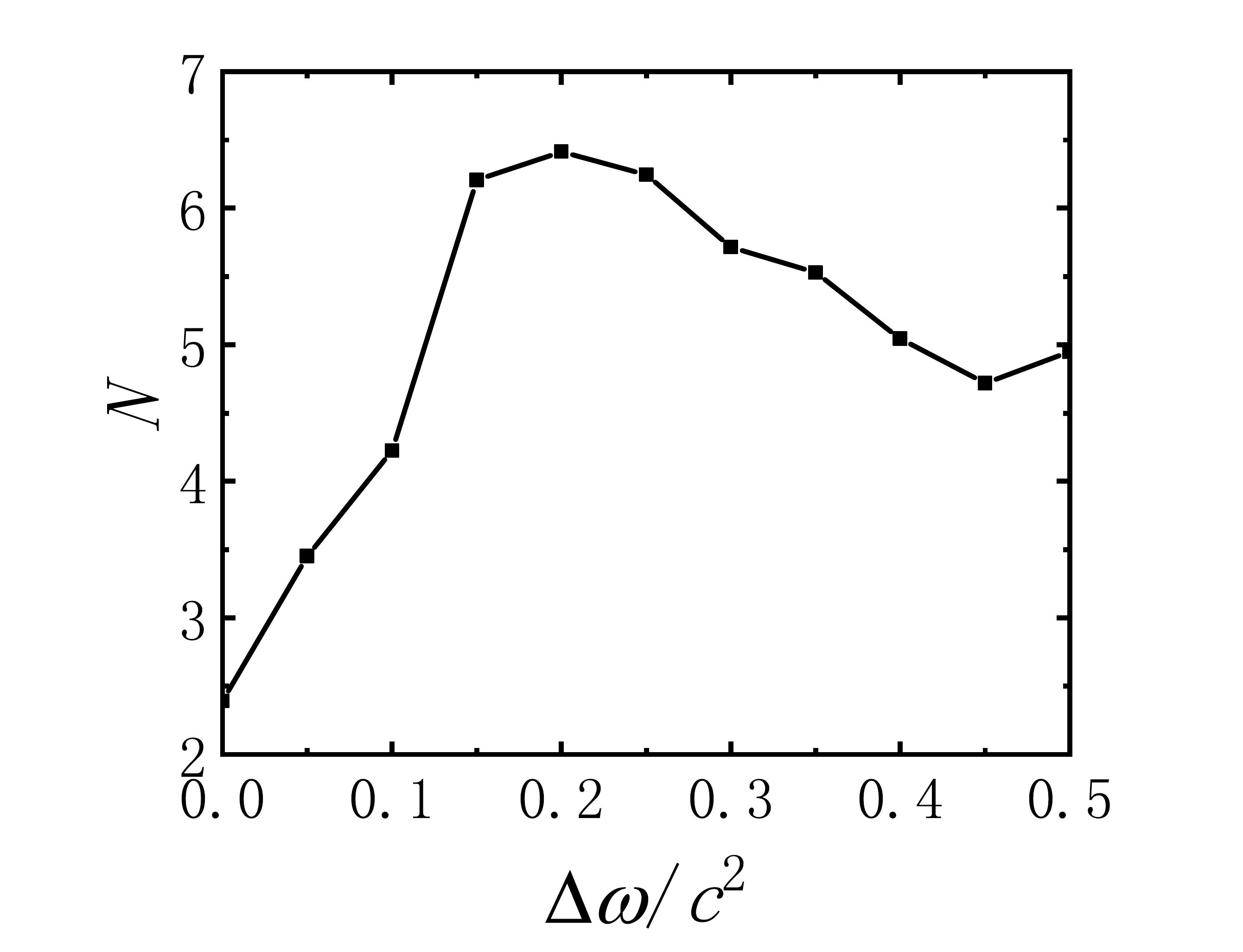}
\caption{\label{fig1} Number of created electrons as a function the FM amplitude $\Delta \omega$. The simulation time is set to $t_1=40\pi/c^2$. Other parameters are $D=10\lambda_C$, $V_1=V_2=1.47c^2$, $\omega_0=0.5c^2$, $\Omega =0.2c^2$. The spatial size is $L=2.0$.}
\end{figure}

In Fig. \ref{fig1}, we represent the number of created electrons as a function of the FM amplitude $\Delta \omega$. Other parameters are set to $D=10\lambda_C$, $V_1=V_2=1.47c^2$, $\omega_0=0.5c^2$, $\Omega =0.2c^2$. The spatial size is $L=2.0$. The simulation time of the numerical calculation is set to $t=40\pi/c^2$, which is less than the time required for the electrons generated in the potential well to leave the simulation regions.
With the increase of $\Delta \omega$, the number of created electrons first increases rapidly, then decreases slowly.
When the FM amplitude is set to $\Delta \omega=0$, the potential well oscillates at a fixed frequency $\omega_0=0.5c^2$, and the number of created electrons is $2.39$. For $\Delta\omega=0.2c^2$, the peak value is $6.42$, which is larger than that of the fixed frequency. It is obviously that chirp effect can promote the generation of electron-positron pairs. The final number of electrons does not increase monotonously with the increase of FM amplitude.

\begin{figure}[htbp]\suppressfloats
\includegraphics[width=14cm]{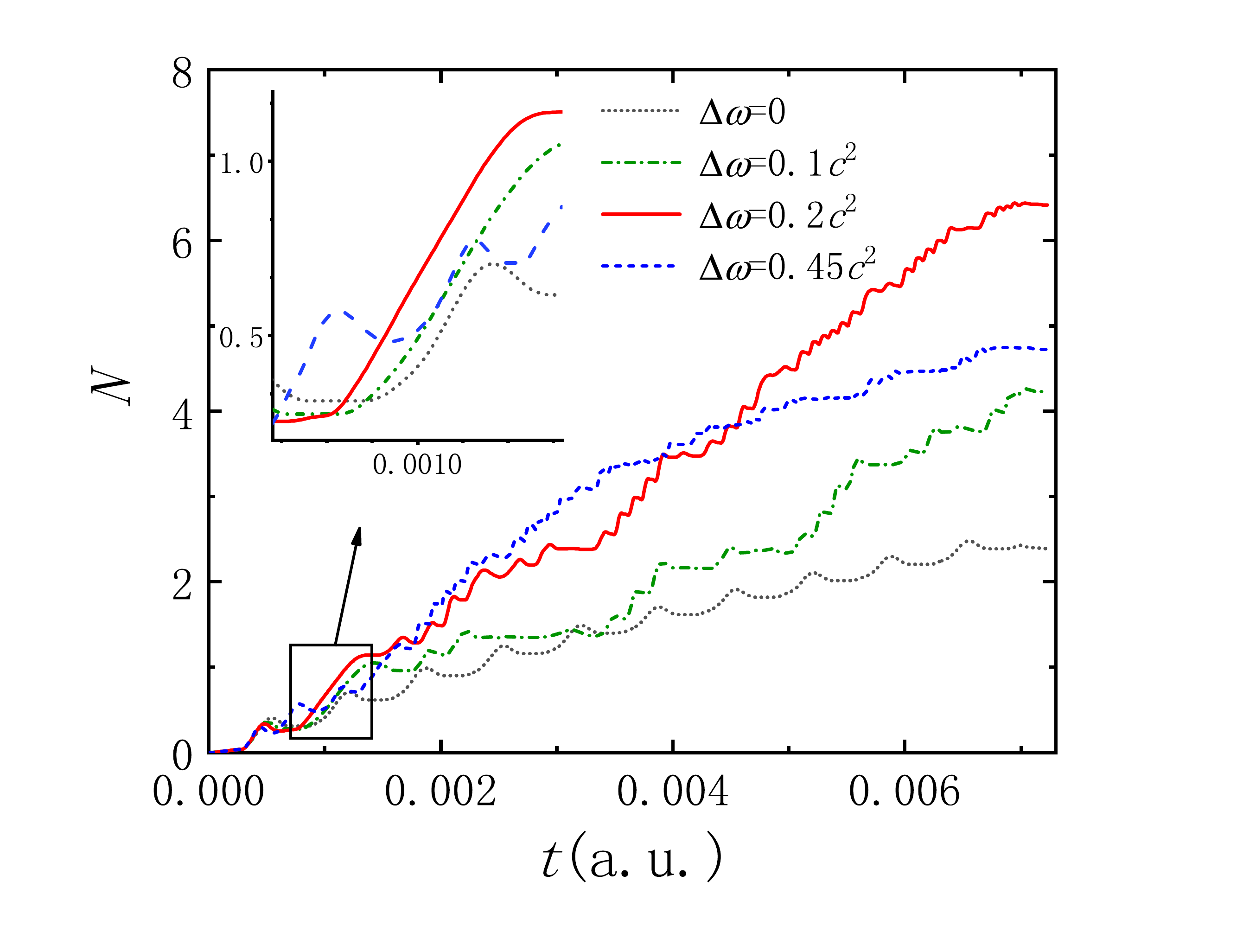}
\caption{\label{fig2} Number of created electrons from $t=0$ to $t=(10+40\pi)/c^2$ for $\Delta \omega=0$ (black dotted line), $\Delta \omega=0.1 c^2$ (green dotted and dashed line), $\Delta \omega=0.2 c^2$ (red solid line) and $\Delta \omega=0.45 c^2$ (blue double-dashed line). Other parameters are same as in Fig. \ref{fig1}. }
\end{figure}

In Fig. \ref{fig2}, the time evolution of the number of created electrons at different FM amplitudes is plotted. For $\Delta \omega =0$, the number grows periodically over time, and the final number is $N=2.39$, which conforms to Fig. \ref{fig1}.
For $\Delta \omega =0.1 c^2$, $0.2 c^2$ and $0.45 c^2$, the number grows unevenly over time.
For $\Delta \omega =0.1 c^2$, the number grows faster than that of the fixed frequency.
For $\Delta \omega =0.2 c^2$, the number growth rate is also larger than that of $\Delta \omega =0.1 c^2$.
However, when the parameter $\Delta \omega$ grows to $0.45c^2$, the number is not always larger than that of $\Delta \omega =0.2 c^2$ with the increase of time. When the evolution time reaches a certain point, the growth rate deceases.

\begin{figure}[htbp]\suppressfloats
\includegraphics[width=16cm]{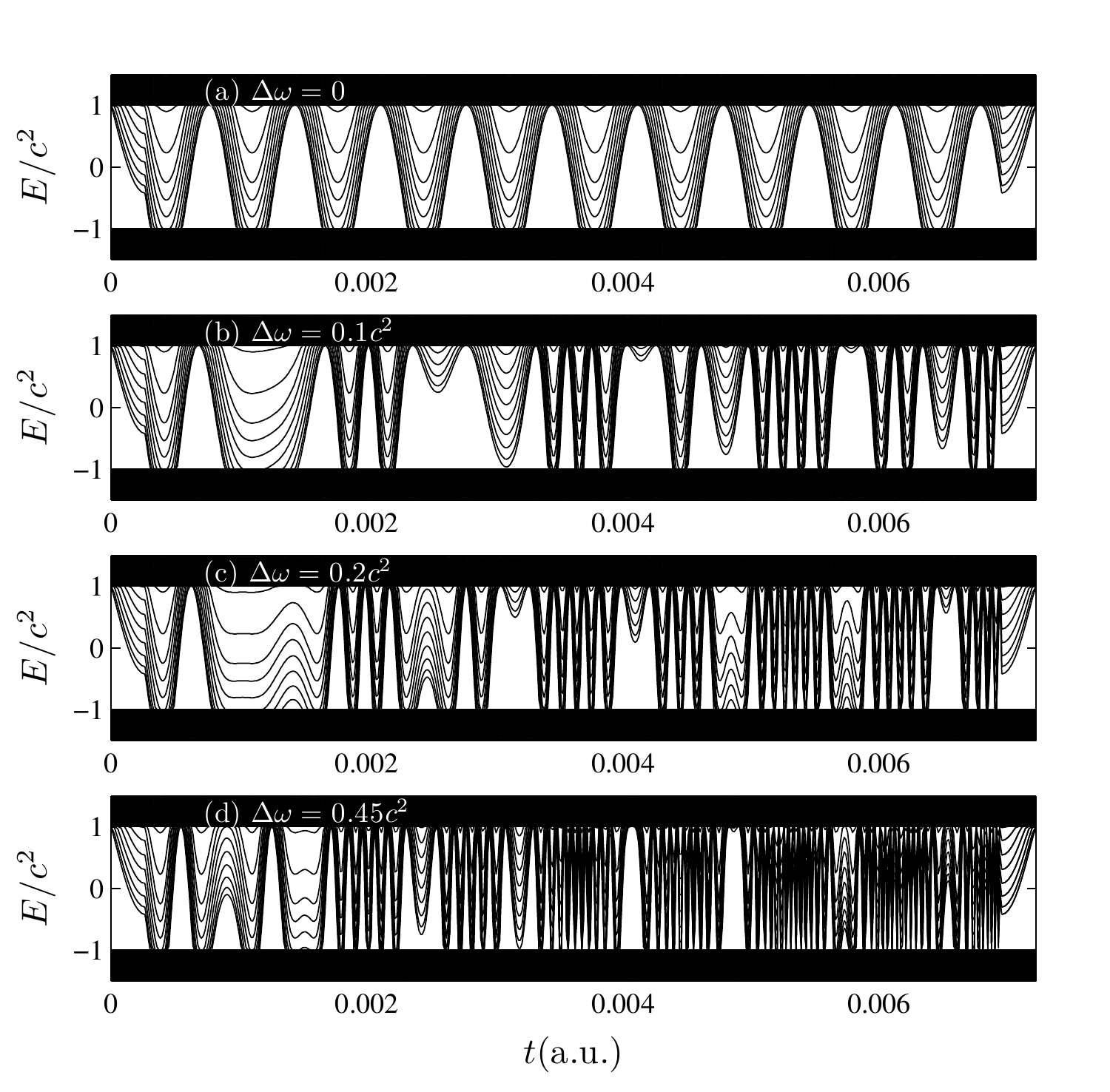}
\caption{\label{fig3} Instantaneous eigenvalues of combined potential wells over time.
Other parameters are the same as those in Fig. \ref{fig2}.}
\end{figure}
To understand above results, we exhibit the instantaneous bound state of combined potential wells for different FM amplitudes in Fig. \ref{fig3}. Other parameters are the same as those in Fig. \ref{fig2}. The evolution of the bound state in processes of turning on and off the static potential is also included in all insets. In Fig. \ref{fig3}(a), when the static potential well is completely turned on, there are eight energy levels in the energy gap. With the increase of time, the bottom five levels periodically dive into the Dirac sea, and then go back to positive continuum states. Compared with the black dotted line in Fig. \ref{fig2}, we find that
the number of electrons grows fastest when the energy level is minimum. This conclusion is consistent with our previous work, in which we introduced the efficient interaction time to explain the production of pairs at a fixed frequency \cite{Wangli2019}. In the simulation time, the energy level dives into the Dirac sea 10 times.

In Figs. \ref{fig3}(b), and (c), with the increase of time, bound states change unevenly, and the efficient interaction time becomes more and more smaller. Near $t=0.001$a.u., the efficient interaction time is lager than that of Fig. \ref{fig3}(a). Compared with the subgraph in Fig. \ref{fig2}, the lager efficient interaction time leads to more electrons created.
In Fig. \ref{fig3}(d), bound states change more rapidly with time. The number of times that bound states dive into the Dirac sea increases with the parameter $\Delta\omega$.

As can be seen above, the final number of electrons is related to the number of times that bound states dive into the Dirac sea, and the efficient interaction time.
Therefore, we propose a function to roughly count the final number of electrons, $N_{final}\varpropto \sum_{i=1}^n r\Delta t$, where $n$ is the number of times that bound states dive into the Dirac sea during the simulation time, $\Delta t$ is the efficient interaction time, $r$ is the growth rate during the efficient interaction time.
According to Fig. 2 in Ref. \cite{Krekora2005}, the number of electrons is quadratic growth when the time set to $t<<1/c^2$, and linear growth for $1/c^2<t<1/c^2+W/c$, where the parameter $W$ corresponds to the width of potential well $D$ in this paper.
In our work, the two parameters that determine the growth rate are about $5.33\times 10^{-5}$a.u. and $5.86\times 10^{-4}$a.u..
In Fig. \ref{fig3}(a), the efficient interaction time of each dive is located between the two parameters, which means the growth rate is constant.
For $\Delta \omega=0.1c^2$, although the efficient interaction time is shortened a little, it also increases the growth rate and the number of dives.
When the modulation amplitude is $0.45c^2$, although the number of dives and the growth rate are very large, the efficient interaction time, especially in the later period of time evolution, is so short that the generation of electrons is suppressed.
Therefore, under the competition mechanism of the growth rate, the number of dives, and the efficient interaction time, there are optimal parameters that make the output the most.
For combined potential wells with $\omega_0=0.5c^2$, the optimal FM amplitude is located at $\Delta\omega_{\rm o}=\Delta\omega_{opt}=0.2c^2$.

\begin{figure}[htbp]\suppressfloats
\includegraphics[width=12cm]{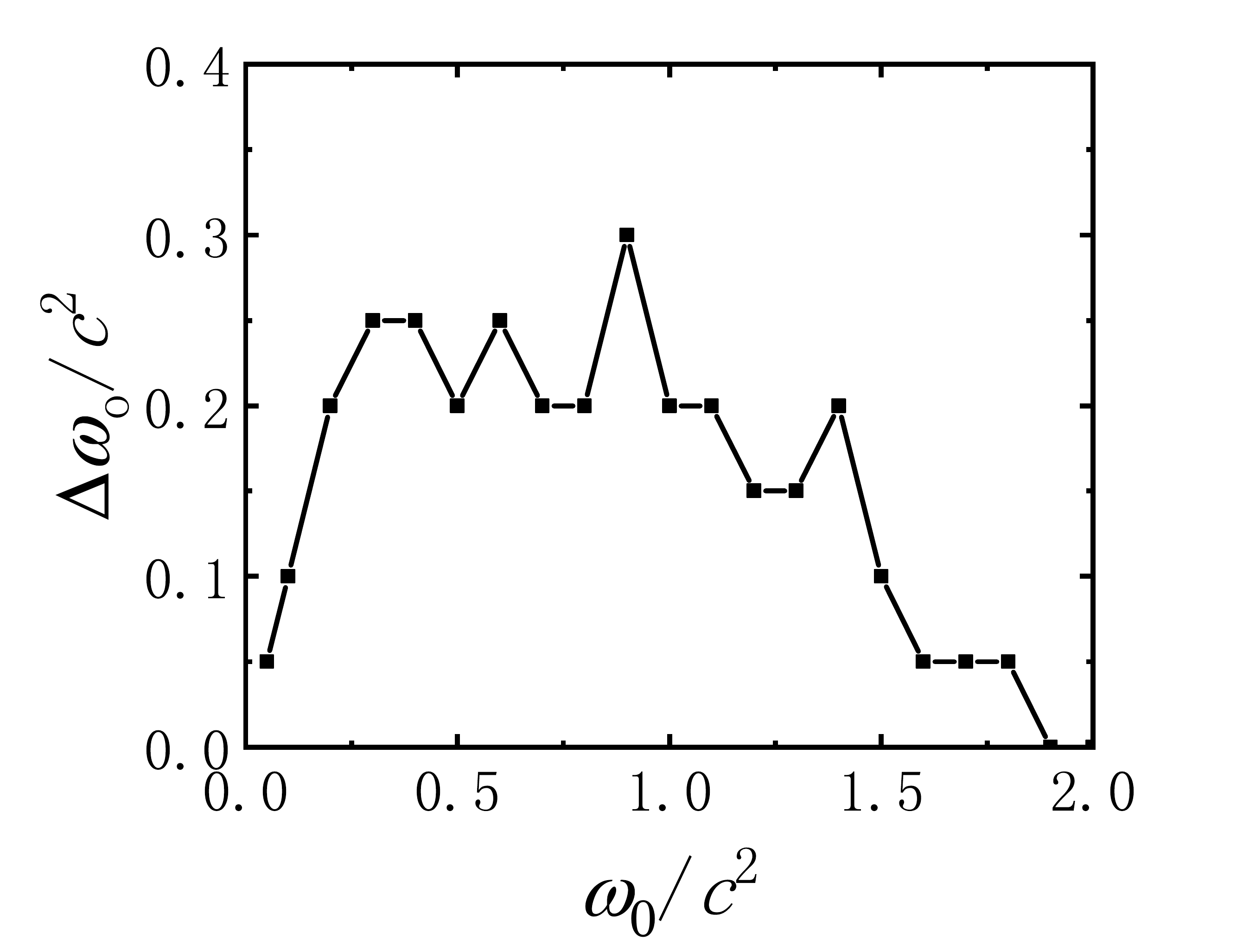}
\caption{\label{fig4} The optimal FM amplitude as a function of the center frequency.
Other parameters are the same as those in Fig. \ref{fig1}.}
\end{figure}

Next, the optimal FM amplitude as a function of the center frequency is shown in Fig. \ref{fig4}.
With the increase of the center frequency, the optimal FM amplitude first increases linearly, and then decays nonlinearly.
For low center frequencies, the parameter $n$ plays a major role in the creation of electrons. The greater the FM amplitude is, the larger the parameter $n$ is.
So, for $\omega_0<0.2c^2$, the optimal FM amplitude satisfies the relation that $\Delta\omega_\texttt{o}=\omega_0$.
For high center frequencies, the generation of electrons is dominated by the efficient interaction time.
Since chirp effect reduces the efficient interaction time, the optimal FM amplitude is smaller for a higher center frequency.

\begin{figure}[htbp]\suppressfloats
\includegraphics[width=12cm]{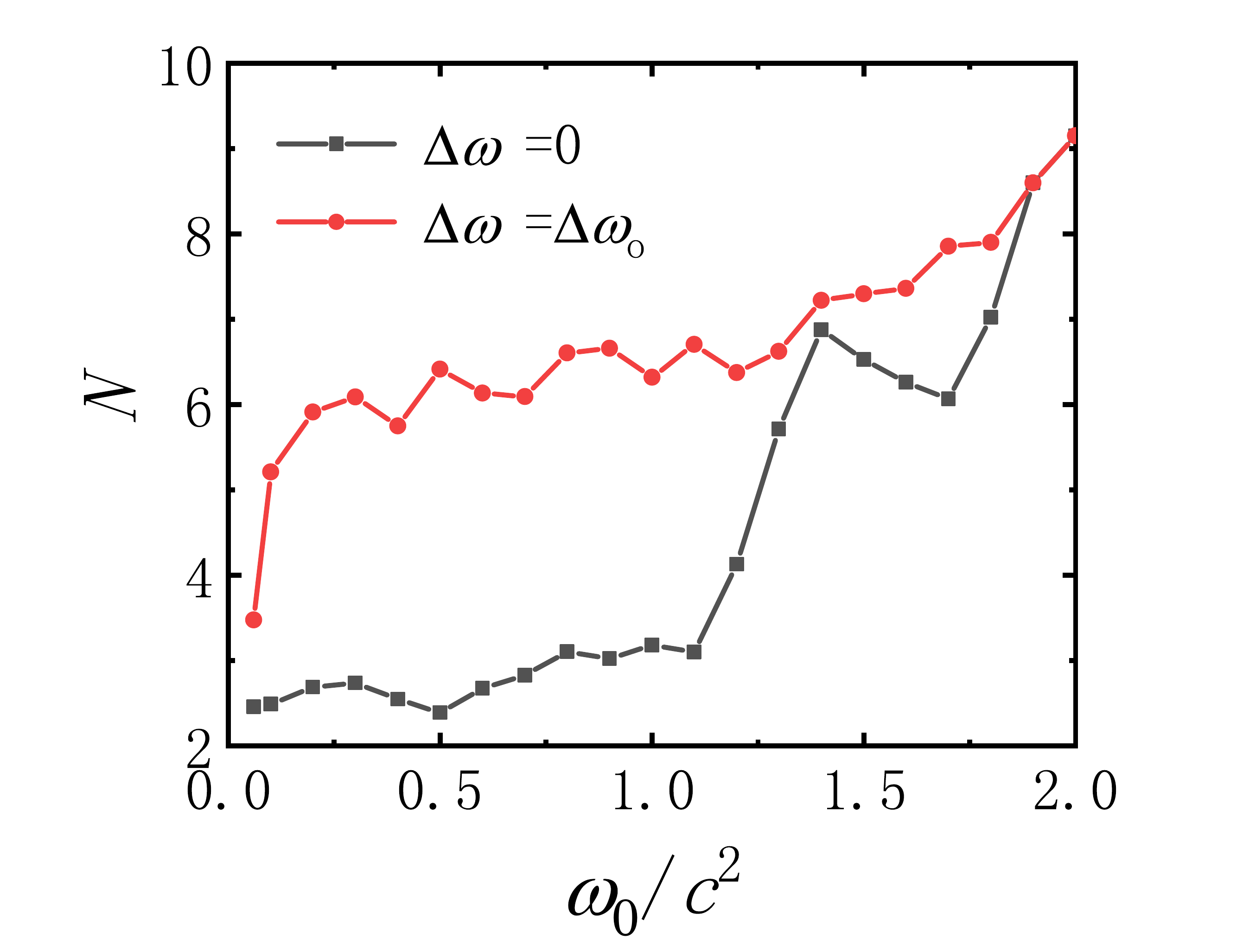}
\caption{\label{fig5} The final number of electrons created in combined potential wells as a function of the center frequency. The black line is for a fixed frequency, and the red line is for the optimal FM amplitude.
Other parameters are the same as those in Fig. \ref{fig1}.}
\end{figure}

In Fig. \ref{fig5}, we represent the final number of electrons created in combined potential wells as a function of the center frequency.
For a fixed frequency potential well, the final number first increases slowly, and then increases quickly with the increase of the center frequency.
When the FM amplitude is optimized, the number of electrons first increases rapidly, and then increases slowly.
Compared with the two lines, it is found that chirp effect has a great influence on the generation of electrons at low frequencies.
For the center frequency that $\omega_0>1.9c^2$, chirp effect does not promote the electron creation.

\begin{table}[htbp]	\label{Table1}
	\centering
	\caption{The center frequency and the ratio of the final number of the optimal FM amplitude to that of the fixed frequency in Fig.\ref{fig5}}
    \renewcommand\arraystretch{1.0}  
    \renewcommand\tabcolsep{10.0pt} 
    \begin{tabular}{l c c c c ccccc ccccc cc}
          \hline \hline
 $\omega_0(c^2)$& 0.06& 0.1& 0.3& 0.5& 0.7& 1.0& 1.2& 1.5& 1.8& 2.0\\
          \hline
          $R$&   1.42& 2.09& 2.22& 2.68& 2.15& 1.99& 1.54& 1.12& 1.12& 1.00\\
          \hline \hline
    \end{tabular}
\end{table}

Table I represents the center frequency and the ratio of the final number of the optimal FM amplitude to that of the fixed frequency in Fig.\ref{fig5}. Taking into account chirp effect, the number of created electrons is increased by about or less than two times. For high frequencies, the enhancement by chirp effect is very small.

\begin{figure}[htbp]\suppressfloats
\includegraphics[width=12cm]{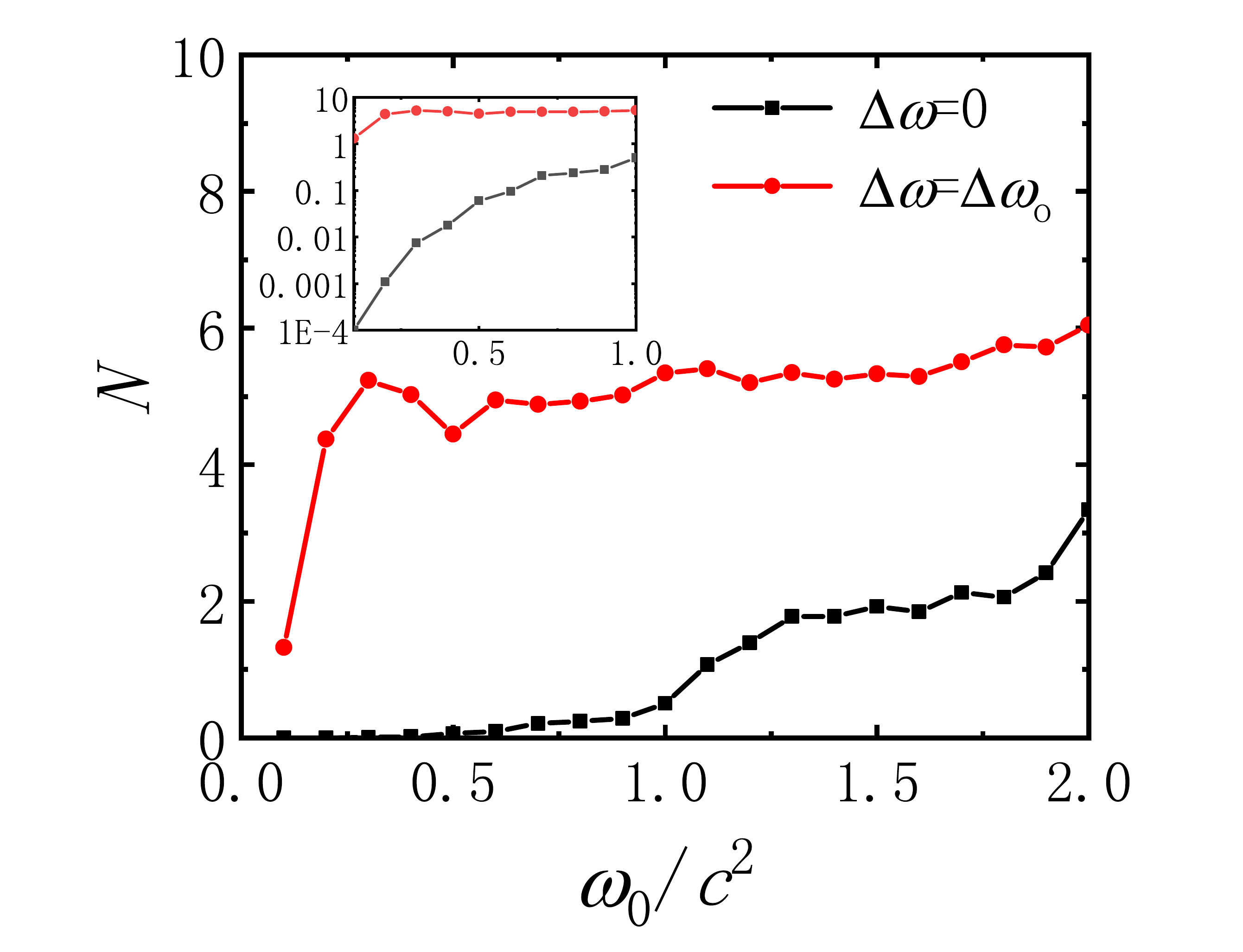}
\caption{\label{fig6} The final number created in a single oscillating potential well as a function of the center frequency.  The black line is for a fixed frequency, and the red line is for the optimal FM amplitude.
Other parameters are the same as those in Fig. \ref{fig1}, except that $V_1=0$.}
\end{figure}

In Fig. \ref{fig6}, we represent the final number created in a single oscillating potential well as a function of the center frequency. For a fixed frequency, the number of created electrons is few at low frequencies, and increases a little at high frequencies.
In the FM potential well, the number of electrons first increases quickly at low frequencies, and then stays around $N=5$, which is larger than that of a fixed frequency.
It is obvious that chirp effect dramatically increases the number of electrons for an oscillating potential well.
In Fig. \ref{fig5}, the number of electrons created in combined potential wells with a fixed frequency is no more than $3$ at low center frequencies, which is smaller than $5$.
By comparison, it is found that for an oscillating potential well with a low fixed frequency, considering chirp effect is better than adding a static potential well.

\begin{table}[htbp]	\label{Table2}
	\centering
	\caption{The center frequency and the ratio of the final number of the optimal FM amplitude to that of a fixed frequency in Fig.\ref{fig6}, where $a(n)$ stands for $a\times 10^n$.}
    \renewcommand\arraystretch{1.0}  
    \renewcommand\tabcolsep{10.0pt} 
    \begin{tabular}{l c c c c ccccc ccccc cc}
          \hline \hline
 $\omega_0(c^2)$&  0.1& 0.2& 0.4&  0.7& 1.0& 1.2& 1.5& 1.8& 2.0\\
          \hline
          $R$&    1.32(4)& 6.98(2)& 2.81(2)&  2.30(1)& 1.10(1)& 3.74& 2.77& 2.80& 1.81\\
          \hline \hline
    \end{tabular}
\end{table}

Table II represents the center frequency and the ratio of the final number of the optimal FM amplitude to that of a fixed frequency in Fig.\ref{fig6}. With the increase of $\omega_0$, the ratio $R$ decreases from a very large number to $1.81$.
It is worth noting that the number of electrons created in the potential well with a fixed frequency is very small, and increases by four orders of magnitude with the increase of $\omega_0$, see the inset of Fig. \ref{fig6}.
Therefore, the number of created electrons increased by up to four orders of magnitude at low center frequencies, and the ratio is much lager than that of table I. Compared with combined potential wells, chirp effect is more sensitive to an oscillating Sauter potential well with low center frequencies.

\begin{figure}[htbp]\suppressfloats
\includegraphics[width=12cm]{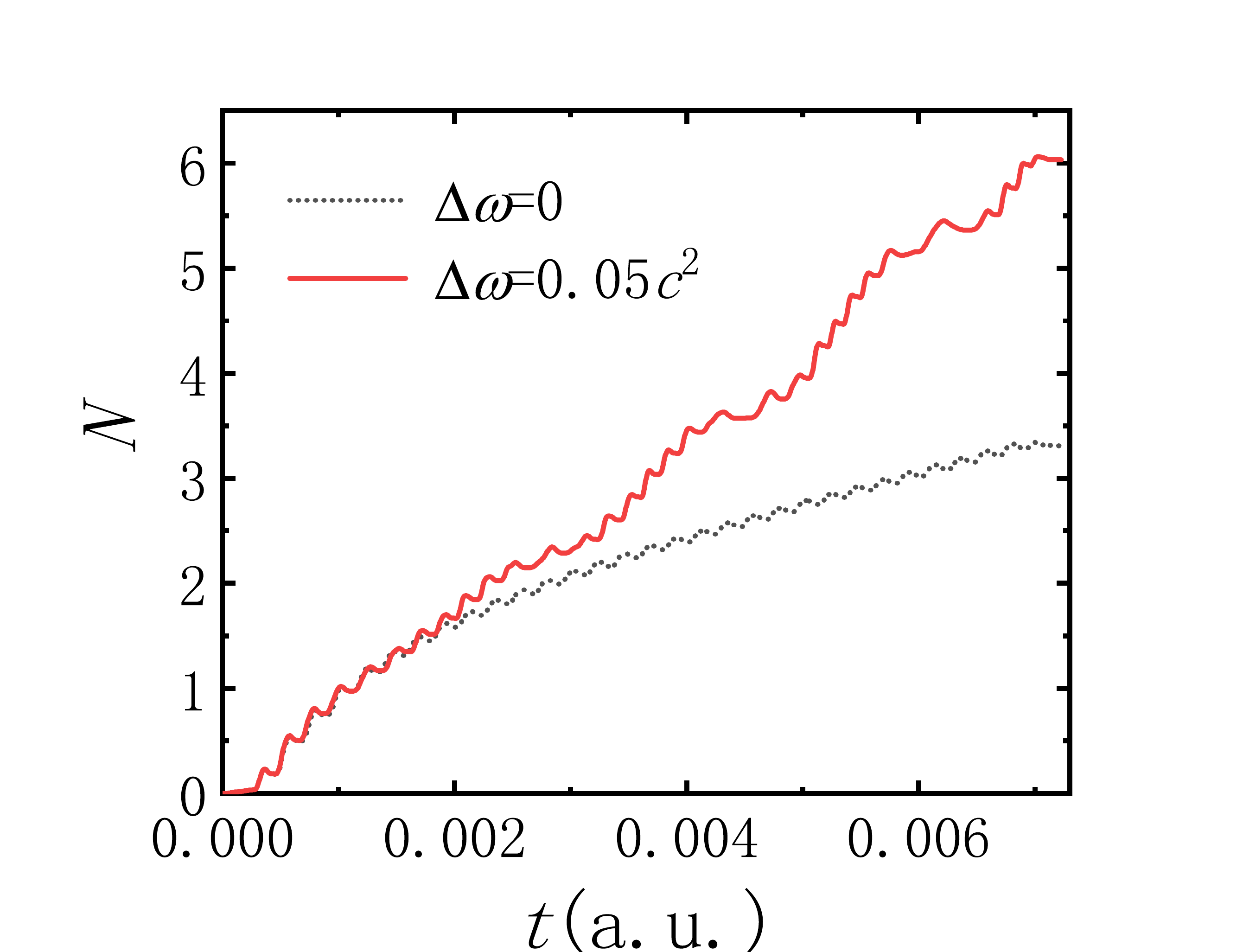}
\caption{\label{fig7} Number of created electrons for $\Delta\omega=0$ (black dotted line) and $0.05c^2$ (red solid line) .
 Other parameters are the same as those in Fig. \ref{fig1}, expect that $\omega_0=1.5c^2$ and $D=4\lambda_C$.}
\end{figure}

For the high center frequency regime, in Fig. \ref{fig7}, the number of created electrons in FM combined potential wells is represented. The center frequency is set to $\omega_0=1.5c^2$, in which the multi-photon process plays an important role in the creation of electrons. In order to study the multi-photon process under chirp effect more easily, we set the width of the combined potential wells to $D=4\lambda_C$. When the frequency is fixed, the number grows slowly over time. After considering chirp effect, the number increases rapidly over time. It is found that chirp effect increases the pair growth rate for a high center frequency.

\begin{figure}[htbp]\suppressfloats
\includegraphics[width=12cm]{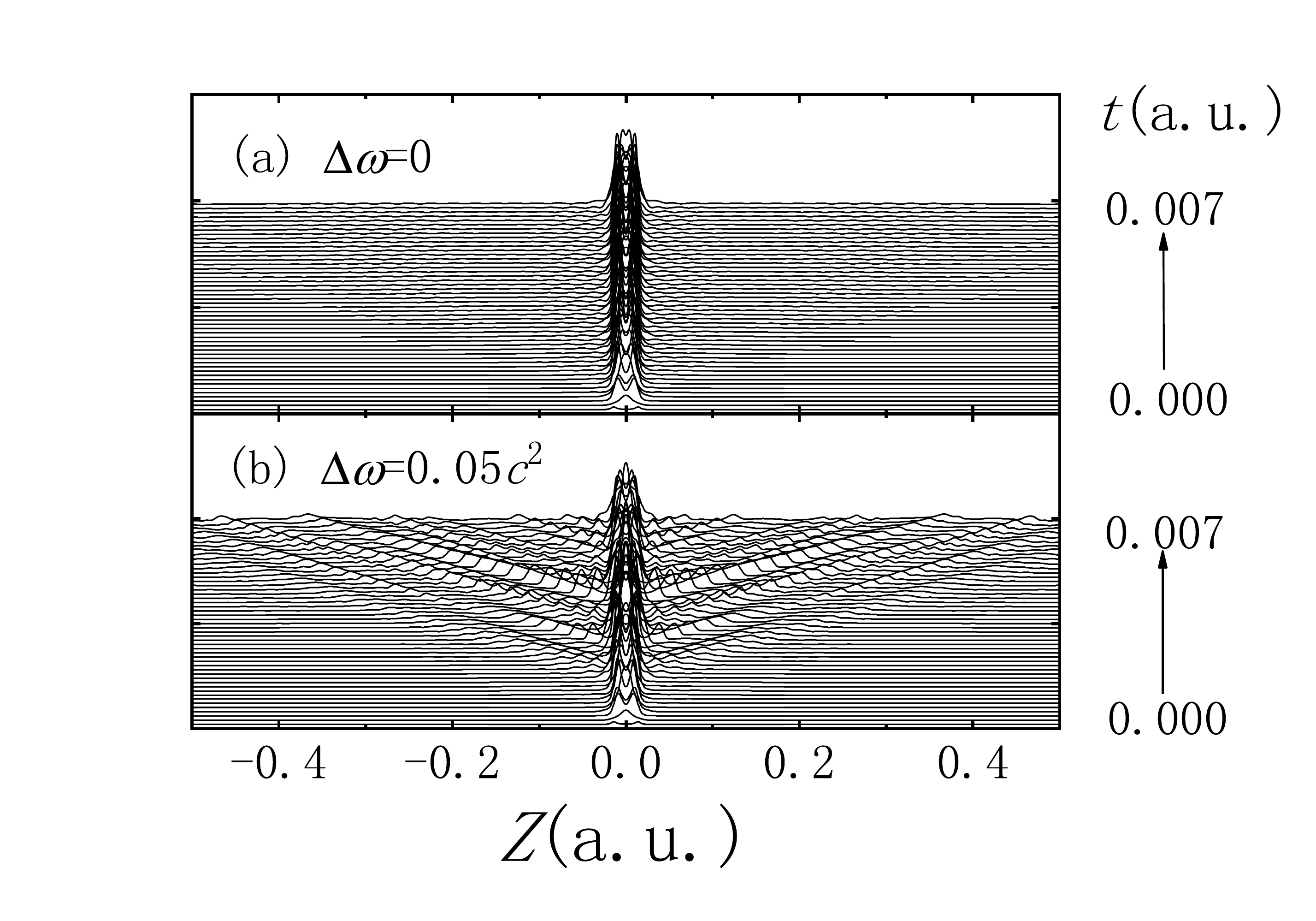}
\caption{\label{fig8} Snapshots of the electronic spatial densities at times $t_k=k\times 1.4\times 10^{-4}$a.u. (k=1,2,3,...,50). Panel (a) is for the fixed frequency potential well with $\omega_0=1.5c^2$, panel (b) is for the FM one with $\omega_0=1.5c^2$ and $\Delta\omega=0.05c^2$. Other parameters are the same as for Fig. \ref{fig7}. }
\end{figure}

In Fig. \ref{fig8}, snapshots of the electronic spatial densities are shown. Panel (a) is for the fixed frequency potential well with $\omega_0=1.5c^2$. Panel (b) is for the FM one with $\omega_0=1.5c^2$ and $\Delta\omega=0.05c^2$.
In Fig. \ref{fig8}(a), created electrons are all trapped in the static potential well during the simulation time.
When chirp effect is taken account, some of electrons escape out of the static potential well, see that in Fig. \ref{fig8}(b).
Therefore, chirp effect reduces Pauli blocking and promotes the creation of electrons.

\begin{figure}[htbp]\suppressfloats
\includegraphics[width=16cm]{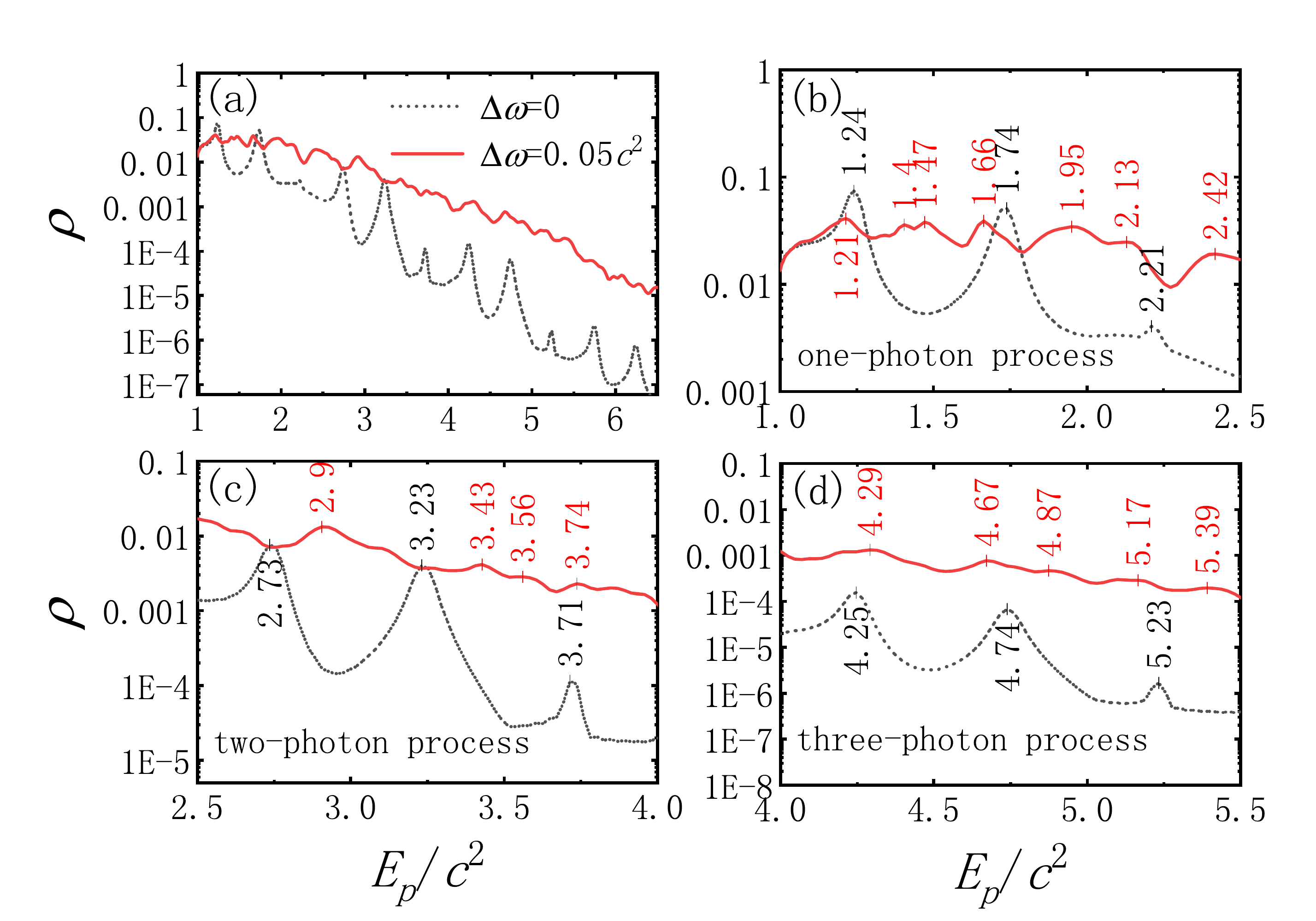}
\caption{\label{fig9} The energy spectrum density of electrons created at the end of the simulation time for $\Delta\omega=0$ (black dotted line) and $0.05c^2$ (red solid line) . Other parameters are the same as those in Fig. \ref{fig7}}
\end{figure}
In Fig. \ref{fig9}, we represent the energy spectrum density of created electrons at the end of the simulation time for $\Delta\omega=0$ and $0.05c^2$. Other parameters are the same as those in Fig. \ref{fig7}. The depth and the width of the static potential well are set to $V_1=1.47c^2$ and $D=4\lambda_C$.
The static potential well provides three energy levels in the gap, which are located at $E_1=-0.2631c^2$, $E_2=0.1762c^2$, and $E_3=0.6698c^2$, respectively.
In Fig. \ref{fig9}(a), the dotted black line is for the fixed frequency with $\omega_0=1.5c^2$.
These discrete peaks correspond to different multi-photon absorption processes.
The red solid line is for the FM potential well with $\Delta\omega=0.05c^2$. The number of peaks is larger, and the value of each peak is marked in Figs. \ref{fig9}(b), (c), and (d).

In Fig. \ref{fig9}(b), we show the one-photon process. For a fixed frequency, these three peaks are located at $E_{P1}=1.24 c^2$, $E_{P2}=1.74c^2$, and $E_{P3}=2.21c^2$, respectively.
These peak values satisfy that $E_{pi}=E_i+\omega_0$ ($i=1,2,3$), which is consistent with that of Ref. \cite{Tang2013}.
For the FM one, the position of the peak moves, and the number of peaks increases. Because the frequency changes from $\omega_0-\Delta\omega$ to $\omega_0+\Delta\omega$, the energy of created electrons also changes from $E_i+(\omega_0-\Delta\omega)$ to $E_i+(\omega_0+\Delta\omega)$. So, new peaks are located between $1.19c^2$ and $1.29c^2$, or $1.69c^2$ and $1.79c^2$, or $2.16c^2$ and $2.26c^2$.
For example, the peak value $1.21c^2$ is between $1.19c^2$ and $1.29c^2$, and corresponds to $E_1+(\omega_0-0.03c^2)$, which means the electron in the first bound state escapes from the energy gap by absorbing one photon with $\omega=1.47c^2$. And the frequency of the photon is located between $\omega_0-\Delta\omega$ and $\omega_0+\Delta\omega$.

\begin{figure}[htbp]\suppressfloats
\includegraphics[width=12cm]{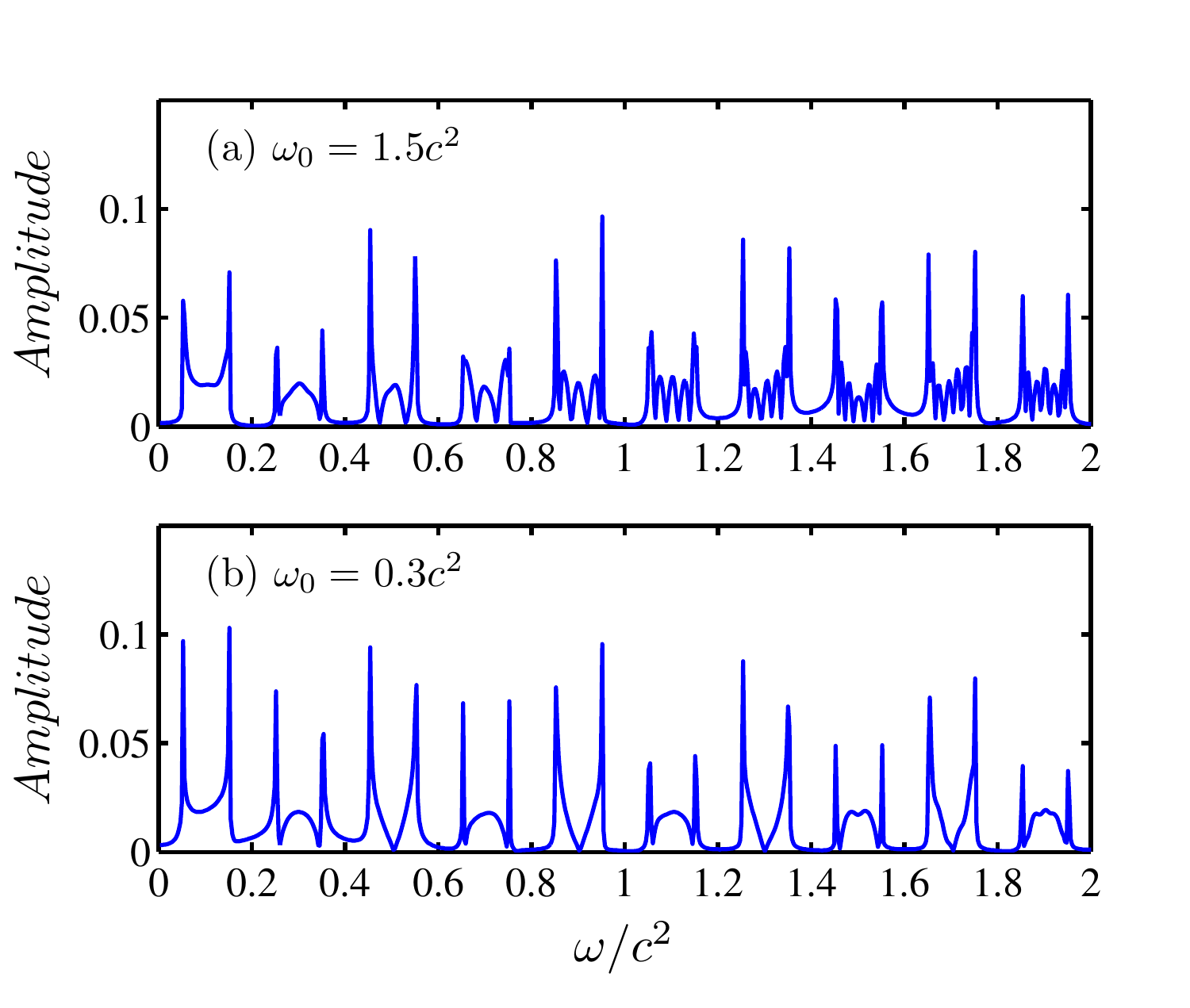}
\caption{\label{fig10} The frequency spectra of the FM potential well with modulation parameters (a) $\omega_0=1.5c^2$, $\Delta\omega=0.05c^2$, (b) $\omega_0=0.3c^2$, $\Delta\omega=0.05c^2$. Other parameters are the same as those in Fig. \ref{fig7}.}
\end{figure}

But some other peaks are not located at these regions, for example the peak value $1.4c^2$.
The reason is that frequencies are more widely distributed in the frequency spectra.
The peak value $1.4c^2$ corresponds to $E_1+(\omega_0+0.16c^2)$,
it means the electron in the first bound state escapes from the energy gap by absorbing one photon with $\omega=1.66c^2$.
Other peak values such as $1.47c^2$, $1.66c^2$, $1.95c^2$, $2.13c^2$, and $2.42c^2$, correspond to $E_1+(\omega_0+0.23c^2)$, $E_1+(\omega_0+0.42c^2)$, $E_2+(\omega_0+0.21c^2)$, $E_2+(\omega_0+0.39c^2)$, and $E_3+(\omega_0+0.21c^2)$, respectively.
These corresponding frequencies of absorbed photons are respectively $1.73c^2$, $1.92c^2$, $1.71c^2$, $1.89c^2$, and $1.71c^2$, which are confirmed in Fig. \ref{fig10}.

In Fig. \ref{fig9}(c) and (d), for a fixed frequency, peak values satisfy $E_{pi}=E_i+n\omega_0$ ($n=2,3$). When chirp effect is considered, the energy of created electrons are located between $E_i+n(\omega_0-\Delta\omega)$ and $E_i+n(\omega_0+\Delta\omega)$.
For multi-photon processes, the energy of electrons is more overlapped than the one-photon process. Therefore, more high-energy electrons are produced under chirp effect.
The growth rate of the number of electrons is enhanced by chirp effect through multi-photon processes.

Now let us to see the frequency spectra of the FM potential well, which is shown in Fig. \ref{fig10}. Panel (a) is to help understand those peaks in Fig. \ref{fig9}. Based on the above analysis, the frequency is not limited between $\omega_0-\Delta\omega$ and $\omega_0+\Delta\omega$, which is seen in Fig. \ref{fig10} (a).
The spectrum is centered at $1.5c^2$, and gradually increases to the left and right within the width of $0.05c^2$. And what is interesting is that it repeats every $0.2c^2$, which is related with the modulation parameter $\Omega=0.2c^2$.
As an illustration, for example, for the peak value $1.66c^2$ of the one-photon process in Fig. \ref{fig9} (b), one can see that it does correspond to $E_1+(\omega_0+0.42c^2)$ but not correspond to $E_2+(\omega_0-0.08c^2)$, since the frequency of $\omega_0+0.42c^2=1.92c^2$ has a higher amplitude than that of $\omega_0-0.08c^2=1.42c^2$ in the frequency spectra. Thus it can be verified that the corresponding frequencies such as $1.73c^2$, $1.92c^2$, $1.71c^2$, and $1.89c^2$ are all revealed in the frequency spectra shown in Fig. \ref{fig10} (a).

Moreover, as a comparison, in Fig. \ref{fig10} (b), the relative low center frequency case is plotted by setting $\omega_0=0.3c^2$.
Strikingly for this low center frequency case, frequency distribution is very similar to that of high center frequency. Since the frequency spectra is not too dependent on the center frequency, the final number of electrons for different center frequencies are almost the same, which explains the result in Fig. \ref{fig6}.

From the results and analysis mentioned above, we can conclude that for a high center frequency, extra low frequencies caused by chirp effect have little effect on the generation of electrons through the quantum tunneling effect, however, for a low center frequency, extra high frequencies caused by chirp effect strongly enhance the generation of electrons through the multi-photon process and the dynamically assisted Schwinger mechanism. That is why chirp effect is more sensitive to the low frequency.

\section{Summary and Conclusion}

We have investigated the chirp effect of FM Sauter potential well on the electron-positron pair creation by employing the computational quantum field theory.
First, we study combined Sauter potential wells with a static one and a FM one to investigate the
number of created electrons and its variation with FM parameters. Second, we remove the static potential well and study chirp effect on the pair creation in a single FM potential well. In order to understand the number changes or/and enhancement, the energy spectrum density and the spatial density of electrons, and the frequency spectra of fields are given as the helpful explanation illustrations.

The main results of our work include:

1. In combined potential wells, chirp effect enhances the number of created electrons. The optimal FM amplitude varies from $0$ to $0.3c^2$. Compared with a fixed frequency, chirp effect can enhance the number of created electrons by about two times.

2. In a single FM oscillating potential well, chirp effect is more sensitive to low center frequencies, and increases the number of created electrons by up to four orders of magnitude compared with the fixed frequency.

3. For an oscillating potential well with a low center frequency, considering chirp effect is better than adding a static potential well.

Chirp effect influences the change of the depth of the potential well with time, causing electrons created in the potential well to escape from the trap, thus promoting the electron pumping. The final number is related to the efficient interaction time, the growth rate, and the number of times that the bound state dives into the Dirac sea. The optimal FM parameters is the result of the competition of these three factors.
Because of chirp effect, the frequency spectrum is widened, and  energies of created electrons change from discrete to continuous, which increases the number of electrons.
Compared with adding a static potential well, considering chirp effect is more conducive to the generation of electrons in a single slowly oscillating potential well.

In this work, the frequency is sinusoidal with time. There are other shape of frequencies that can be considered and studied. The innovation of this paper is to introduce the relation of the final number to the efficient interaction time, the growth rate, and the number of times that the bound state dives into the Dirac sea, which can well explain the generation process of electrons. It is significant to conclude that chirp effect is more effective at low center frequencies.

\begin{acknowledgments}
This work was supported by the National Natural Science Foundation of China (NSFC) under Grants No. 11875007 and No. 11935008. The computation was carried out at the HSCC of the Beijing Normal University.
\end{acknowledgments}

\end{document}